\documentclass[]{spie}  

\usepackage{amsmath,amsfonts,amssymb}
\usepackage{graphicx}
\usepackage[colorlinks=true, allcolors=blue]{hyperref}
\usepackage[separate-uncertainty=true,range-phrase=--]{siunitx}

\title{Beam combiner for the Asgard/BIFROST instrument}

\author[a]{Daniel J. Mortimer}
\author[a]{Sorabh Chhabra}
\author[a]{Stefan Kraus}
\author[b]{Narsireddy Anugu}
\author[c]{Romain Laugier}
\author[d]{Jean-Baptiste Le Bouquin}
\author[e]{John D. Monnier}

\affil[a]{University of Exeter, School of Physics and Astronomy, Stocker Road, Exeter, EX2 7SJ, UK}
\affil[b]{CHARA Array, Mt. Wilson Observatory, Mt. Wilson, California, USA}
\affil[c]{Institute of Astronomy, KU Leuven, Celestijnenlaan 200D, 3001 Leuven, Belgium}
\affil[d]{Institut de Plan\'etologie et d’Astrophysique de Grenoble, UMR 5274, Grenoble, France}
\affil[e]{Department of Astronomy, University of Michigan, 500 Church Street, Ann Arbor, MI 48109, USA}

\authorinfo{Further author information: Send correspondence to D.J.M.\\ E-mail: d.j.mortimer@exeter.ac.uk}

\begin{document} 
\maketitle

\begin{abstract}

BIFROST will be a short-wavelength ($\lambda$~=~1.0~-~\SI{1.7}{\micro\meter}) beam combiner for the VLT Interferometer, combining both high spatial ($\lambda$/2B~=~0.8 mas) and spectral (up to R~=~25,000) resolution. It will be part of the Asgard Suite of visitor instruments. The new window of high spectral resolution, short wavelength observations brings with it new challenges. Here we outline the instrumental design of BIFROST, highlighting which beam combiner subsystems are required and why. This is followed by a comparison All-In-One (AIO) beam combination scheme and an Integrated Optics (IO) scheme with ABCD modulation both in terms of expected sensitivity and the practical implementation of each system. 

\end{abstract}


\section{Introduction}
\label{sec:intro}  

BIFROST is a planned visitor instrument for the Very Large Telescope Interferometer (VLTI) and part of the ASGARD suite of instruments (Martinod et al. 2022)\cite{asgard_martinod}. With its short wavelength (J band) and high spectral resolution (R~=~25,000) will open up a new observational window at the VLTI. A few key science drivers include measuring precision dynamical masses and ages for GAIA binaries, measuring the spin-orbit alignment for directly imaged exoplanets and off-axis exoplanet spectroscopy (Kraus et al. 2022)\cite{bifrost_kraus}.

The instrument itself is still in the design phase, a key part of which is identifying which subsystems are necessary for the instrument to function. In addition to this the implications of the different beam combination schemes, namely an All-In-One (AIO) or a Integrated Optics (IO) beam combiner must be considered. 

In this proceeding we detail our design considerations to date, outlining the need for an Atmospheric Dispersion Corrector (ADC), Longitudinal Dispersion Corrector (LDC), Lithium Niobate Plates (LNP) and a fiber injection unit. This is followed by a discussion on the trade-offs between an All-In-One (AIO) and Integrated Optics (IO) beam combination system. 

\section{Atmospheric dispersion corrector} \label{sec:adc}

The variation of refractive index of the atmosphere as a function of wavelength gives rise to angular dispersion. This is the effect where the observed position of an object on-sky changes as a function of wavelength. For an instrument such as BIFROST which couples light through single mode fibers, to first order its effective field of view on-sky is the diffraction limited psf of the telescope, which is around 70 mas at $\lambda$~=~\SI{1.1}{\micro\meter} on the Unit Telescopes (UTs). If the light across one band is dispersed over an angular range larger than the field of view of the fiber then not all wavelengths will be coupled into the fiber simultaneously. In which case an ADC will be required to remove the atmospheric angular dispersion. 

\begin{figure}
    \centering
    \includegraphics[width=0.75\columnwidth]{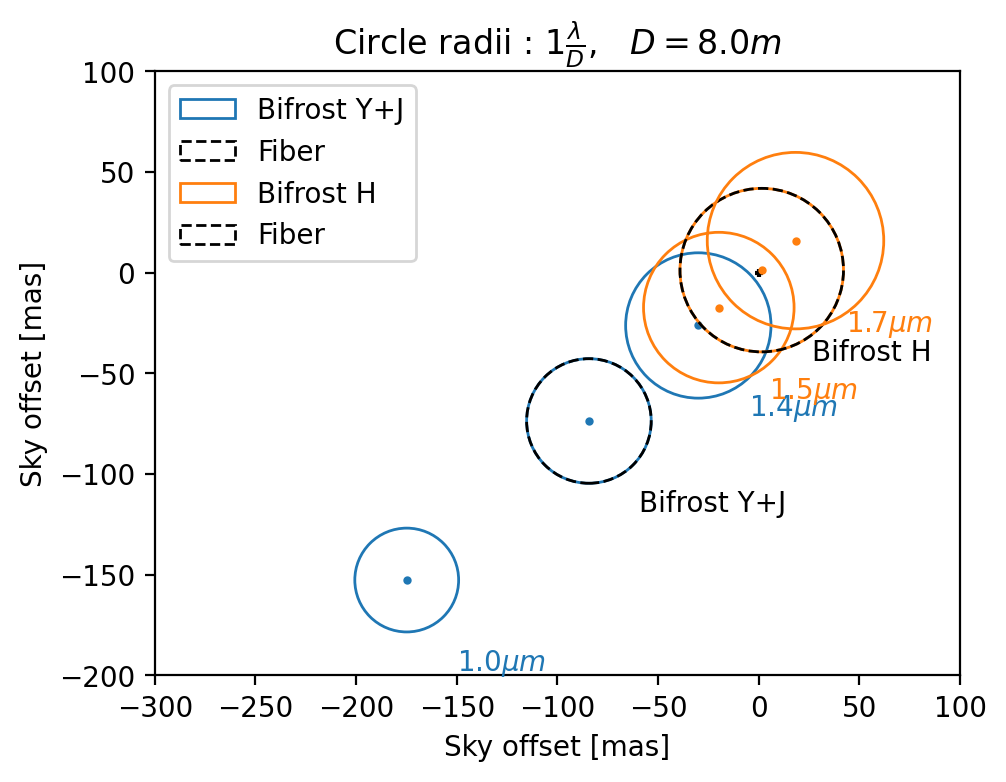}
    \caption{The on-sky offsets for the centres and edges of the Y+J and H band modes of BIFROST. This plot represents the reasonable worst case scenario, observations on the UTs at an altitude of 30$^{\circ}$ with no ADC present. The diameter of the circles represent the diffraction limited field of view of the UTs at the wavelength shown. The black dashed lines represent where the single mode fiber tips would be placed to maximise coupling at the centre of each band. In the H band a significant fraction of the light at the edges of the band are not coupled into the single mode fiber. However the problem is more pronounced in the Y+J band where the edges of the band miss the fiber tip completely. To remove this on-sky offset an ADC will be necessary.}
    \label{fig:BIFROST_ADC_requirement}
\end{figure}

To determine if an ADC is necessary for BIFROST we used the atmospheric model of SCIFYsim (Laugier et al. 2021)\cite{2021sf2a.conf..339L} to estimate the on-sky offset for light at the edges with respect to the centres of our bands for the reasonable worst case scenario, an object at an altitude of 30$^{\circ}$ observed on the UTs. The results are shown in figure~\ref{fig:BIFROST_ADC_requirement}. 

Figure~\ref{fig:BIFROST_ADC_requirement} shows the position of the centres and edges of our Y+J (\SI{1.0}{} - \SI{1.4}{\micro\meter}) and H (\SI{1.5}{} - \SI{1.7}{\micro\meter}) band modes. The circles represent the approximate size of the diffraction limited PSFs for the various wavelengths. The black dashed rings are the effective field of view of the fiber which is placed on-top of the central wavelength of the two bands. In the H band when the fiber is positioned at the centre of the band, light from the edges does still fall within the field of view of the fiber. The edges of the band are displaced by 60\% of the radius of the fibers field of view. Approximating the beam profile that couples into the fiber as a Gaussian, the beam intensity is given by 

\begin{equation}
    I(r) = I(0) \textrm{exp}\left(\dfrac{-2r^2}{\omega(0)^2}\right),
\end{equation}

where $I(0)$ is the intensity at the peak, $r$ the distance from the centre of the beam and $\omega(0)$ the Gaussian beam radius (the radius at which the intensity has decreased to $1/e^2$). Therefore, for a beam displaced by 60\% of the fiber radius (the fiber radius being taken as $\omega(0)$) the intensity of the beam at the core of the fiber is only 50\% of the peak of the beam, leading to significant coupling losses. 

For the Y+J mode the result is even more significant, as figure~\ref{fig:BIFROST_ADC_requirement} shows, the field of view of the fiber placed at the centre of the band does not overlap at all with the edges of the band, leading to zero flux being coupled into the fiber from wavelengths at the edges of the band. 

Given these results an ADC will be required for BIFROST when observing with the UTs. Due to the smaller telescope diameters of \SI{1.8}{\meter} on the Auxiliary Telescopes (ATs) the fiber field of view will be larger and hence may not require an ADC. Observations at higher altitudes will also have looser constraints on the need for an ADC. 

One option would be to design an ADC that can be inserted and removed from the beam path such that it can be used for pointings at low altitudes, especially on the UTs and removed for observations where the loss of throughput due to the glass of the ADC exceeds the gain in coupling efficiency into the optical fibers. 

\section{Longitudinal Dispersion Corrector} \label{Sec:LDC}

Observations of targets not at zenith require a different path length through the delay lines for the different beams of starlight being combined to compensate for the geometric path delay. As the refractive index of air varies as a function of wavelength, if the optical path length of two beam lines is equalised at one wavelength (the fringe tracking wavelength) there will be a residual Optical Path Difference (OPD) error between the beam lines at other wavelengths. 

To demonstrate this we have built a model to estimate the visibility loss as a function of the difference in air path between the beam lines. As the difference in air path increases, so does the residual OPD error which will push the interference fringes away from the centre of the coherence envelope, reducing the observed fringe contrast, in the worst case scenario removing the interference fringes entirely. A Longitudinal Dispersion Compensator (LDC) is designed to counteract the differential optical path length in air.

The length of the coherence envelope is given by $L = \lambda^2/\Delta\lambda$ where $\lambda$ is the central wavelength of observation and $\Delta\lambda$ the range of wavelengths being interfered. Therefore an LDC is more likely to be needed at lower spectral resolutions where each spectral channel is more broadband. Haniff (2007)\cite{2007NewAR..51..565H} gives the maximum visibility due to the coherence envelope to be 

\begin{equation}
    V = sinc\left(\dfrac{\pi D\Delta\lambda}{\lambda^2}\right),
\end{equation}
where $D$ is the OPD between the two beams being interfered. To establish if an LDC is required for BIFROST we plot the visibility as a function of the difference in air path between the two beam lines.

The OPD is given by $\textrm{OPD} = l(n_{\textrm{obs}} - n_{\textrm{ft}})$ where $n_{\textrm{obs}}$ is the refractive index of air at the wavelength of observation and $n_{\textrm{ft}}$ the fringe tracking wavelength. For wavelengths in the range $\lambda$ = \SI{0.3}{} - \SI{1.69}{\micro\meter} we use the model of Ciddor (1996)\cite{1996ApOpt..35.1566C} and for $\lambda$ = \SI{1.7}{} - \SI{2.5}{\micro\meter} Mathar (2007)\cite{2007JOptA...9..470M}. Ambient conditions of temperature 15$^\circ$C and a pressure of 1014 mbar are assumed. The wavelength range on a spectral channel, $\Delta\lambda$ is estimated from the spectral resolution R value. 

\begin{figure}
    \centering
    \includegraphics[width=0.8\columnwidth]{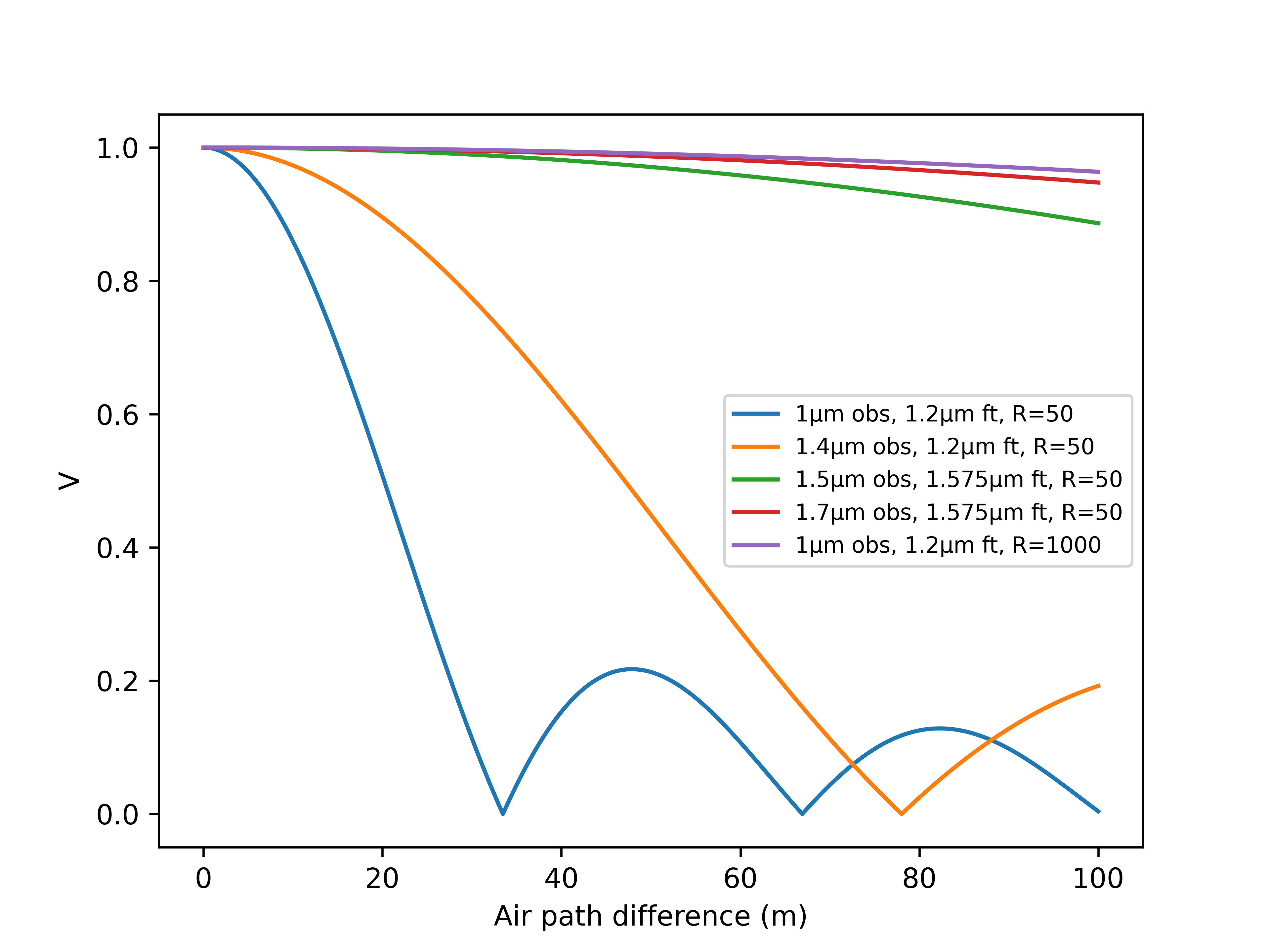}
    \caption{The visibility loss as a function of the difference in air path length between two beams lines being combined. Here the OPD error between the beam lines is assumed to be zero at the centres of the Y+J and H band modes. The resulting drop in visibility as the OPD error grows with larger differential path lengths is plotted for the edges of the bands for both the R~=~50 and R~=~1,000 modes of BIFROST. The loss in the R~=~1,000 mode is shown to be negligible however the losses in the R~=~50 mode can be significant, especially in the Y+J band where the first null of the coherence envelope is reached after \SI{35}{\meter} of differential air path.}
    \label{fig:no_LDC_vis_loss}
\end{figure}

While fringe tracking for Asgard is performed in the K band it is possible to use the differential delay lines of BIFROST to change the wavelength which has zero OPD error by applying a static offset from the main delay lines. Therefore we examine the fringe contrast loss at the edges of the bands for the two modes of BIFROST (Y+J, $\lambda$ = \SI{1}{} - \SI{1.4}{\micro\meter} and H, $\lambda$ = \SI{1.5}{} - \SI{1.7}{\micro\meter}) when the OPD error is set to zero at the centre of each band. The results are shown in figure~\ref{fig:no_LDC_vis_loss}. As the figure shows the visibility losses for the R~=~1,000 spectral mode are negligible, the loss in throughput from the extra glass of the LDC will likely result in a worse SNR than if no LDC is placed in the beam path. The results are more ambiguous for the lowest spectral mode planned for BIFROST, R~=~50. In the H band visibilities drop to 88\% for differential air paths of \SI{100}{\meter}. The effect is even more pronounced in the J band where the fringe contrast drops to zero (the first null of the coherence envelope) for an air path of \SI{35}{\meter}. Such a large loss in visibility in the low resolution arm would make it unusable in the J band making an LDC necessary for BIFROST. 

Now that the need for an LDC has been established its location in the beam train must be considered. The first option is to put it in the BIFROST only part of the Asgard beam train, this will maximise the throughput of the fringe tracker Heimdallr. However, will lead to non-common path OPD errors between Heimdallr and BIFROST as the thickness of the LDCs are adjusted. This could in theory be compensated by modelling the change in OPD as the LDC glass thickness is adjusted and correcting this with the DDLs of BIFROST but placing the LDC in the common path of Heimdallr and BIFROST would be the simpler implementation.

\section{Lithium Niobate plates} \label{sec:Lithium_niobate_plates}

The use of optical fibers can reduce the instrumental contrast if orthogonal polarisation states have unmatched phases. To minimise this effect we will follow the design of MIRC-X (Anugu et al. 2020)\cite{2020AJ....160..158A} and PIONIER (Lazareff et al. 2012)\cite{2012A&A...543A..31L} and place Lithium Niobate (LiNbO3) plates in the beam path which can be rotated to introduce a differential phase shift between the polarisation states. 

\begin{figure}
    \centering
    \includegraphics[width=0.6\columnwidth]{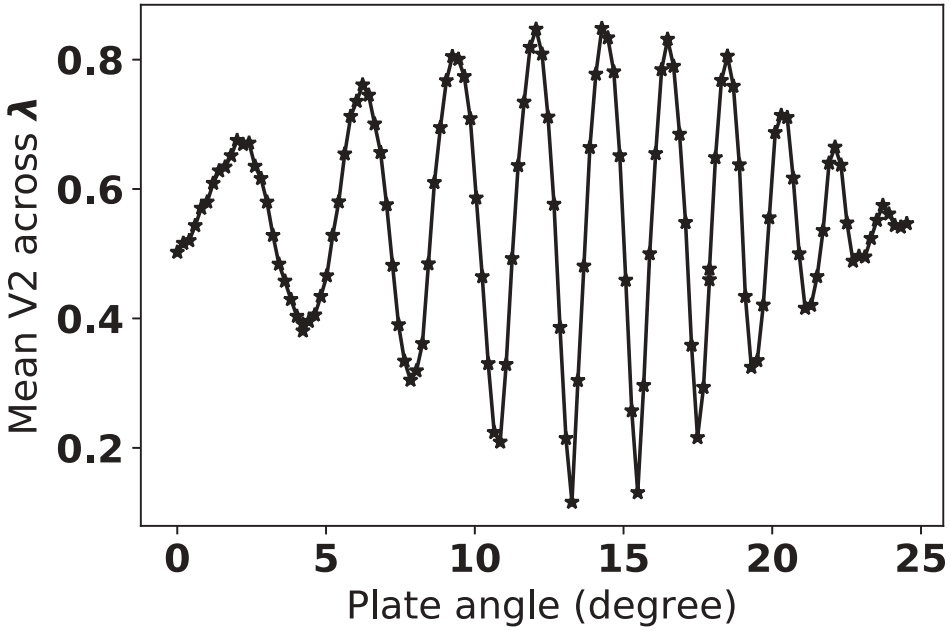}
    \caption{Laboratory data of the instrumental fringe contrast as a function of the LNP rotation angle for the MIRC-X instrument. The graph shows the increase in instrumental contrast as the two orthogonal polarisation states are phased up with each other. This is of Figure 4 from~Anugu et al. (2020)\cite{2020AJ....160..158A}, \copyright~AAS. Reproduced with permission.}
    \label{fig:MIRC_X_LNB}
\end{figure}

An example of the effects of these plates in the MIRC-X instrument is shown in figure~\ref{fig:MIRC_X_LNB} which shows the significant increase in instrumental visibility as the phases of the two polarisation states are equalised. 

\section{Fiber injection unit}

One the the biggest technical risks to the development of BIFROST is maintaining a good injection into the single mode fibers owing to the tight alignment tolerances involved in coupling atmospherically perturbed starlight into a single mode fiber with an $\sim$~\SI{6}{\micro\meter} core. The primary role of the fiber injection unit is to take the collimated beams of starlight and couple them into the fibers however it will also act as the Differential Delay Lines (DDLs) of BIFROST to enable us to cophase BIFROST with the other instruments on the Asgard table, as well as allowing us to apply global offsets to the path length as discussed in section~\ref{Sec:LDC}. 

On the Asgard table at the VLTI we will have one Deformable Mirror (DM) per beam line controlled by the Baldr instrument to correct any wavefront errors in the beams reaching the Asgard table. This should mean that once aligned with the beams coming from the DMs, the fiber injection into BIFROST will be stable, with only long term instrumental drifts of the optics between the DMs and the BIFROST fibers affecting the coupling. However we must have a way to correct for these instrumental drifts, either by steering the beams towards the fibers, as is done in the MYSTIC combiner (Monnier et al. 2018)\cite{2018SPIE10701E..22M}, or moving the fiber tips to the beams as is done in the MIRC-X combiner (Anugu et al. 2020)\cite{2020AJ....160..158A}.

\begin{figure}
    \centering
    \includegraphics[width=0.9\columnwidth]{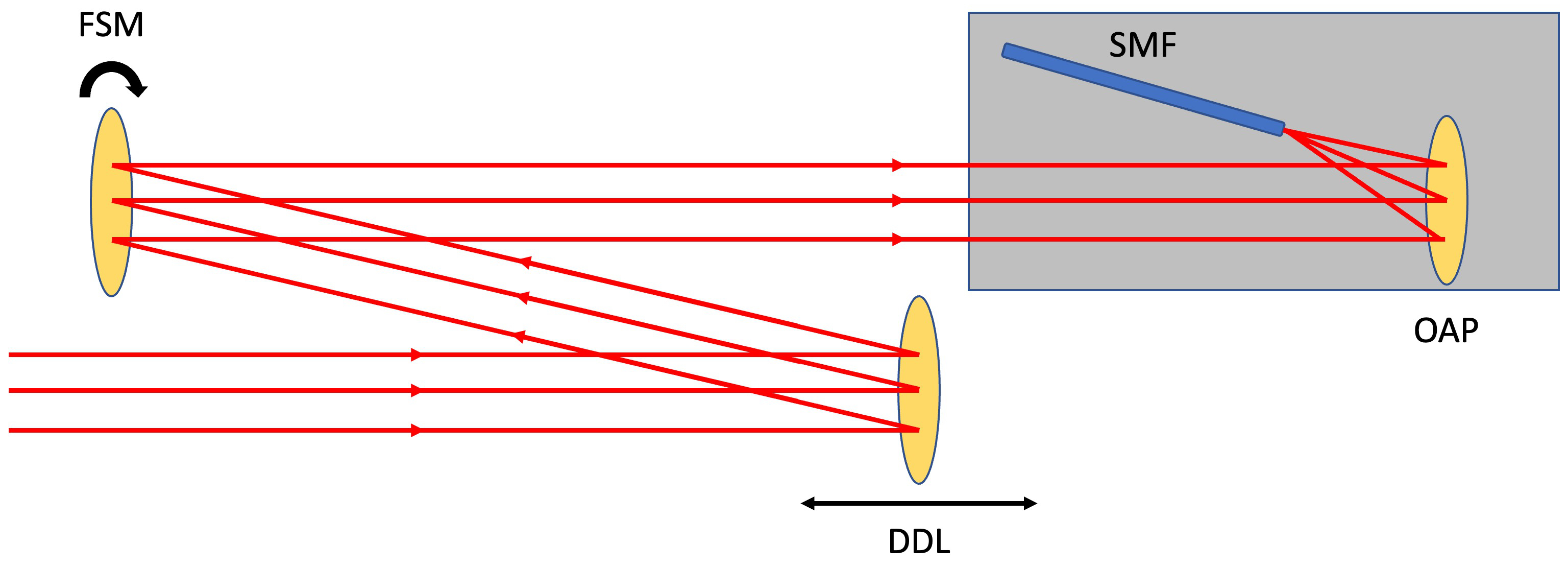}
    \caption{One potential layout of the BIFROST fiber injection unit. Here the beams first reflects off a dedicated DDL which adjusts the path length of the beams before reflecting off a FSM that tips and tilts to optimise injection into the SMF. In this scheme the OAP and SFM remain static, reducing the risk of them becoming misaligned. This diagram is Not to scale.}
    \label{fig:MYSTIC_style_injection}
\end{figure}

The layout of one design we are considering is shown in figure~\ref{fig:MYSTIC_style_injection} and is the same layout as implemented by MYSTIC. Here the first optic is the DDL, a flat mirror that is motorised to move along one axis to add or remove optical path for a single beam. This is followed by a Fast Steering Mirror (FSM) which allows for tip-tilt corrections to steer the beam and optimise injection into the fiber. Finally there is the Off-Axis-Parabola (OAP) and Single Mode Fiber (SMF) unit which are mounted together to maintain the optical alignment between the two. 

The advantage of this scheme is that the OAP and SMF do not move. This ensures both that the optical alignment of the OAP and SMF (which have very tight alignment tolerances with respect to each other, see Mortimer (2021)\cite{Mortimer_PhD_thesis}) will not become misaligned due to vibrations or non-repeatability of moving mounts. The disadvantage of such a scheme is that this introduces an extra two reflections (the DDL and FSM). Taking a typical reflection loss of a protected gold coated mirror of 2.5\% in the infrared this gives a 5\% reduction in throughput.

\begin{figure}
    \centering
    \includegraphics[width=0.75\columnwidth]{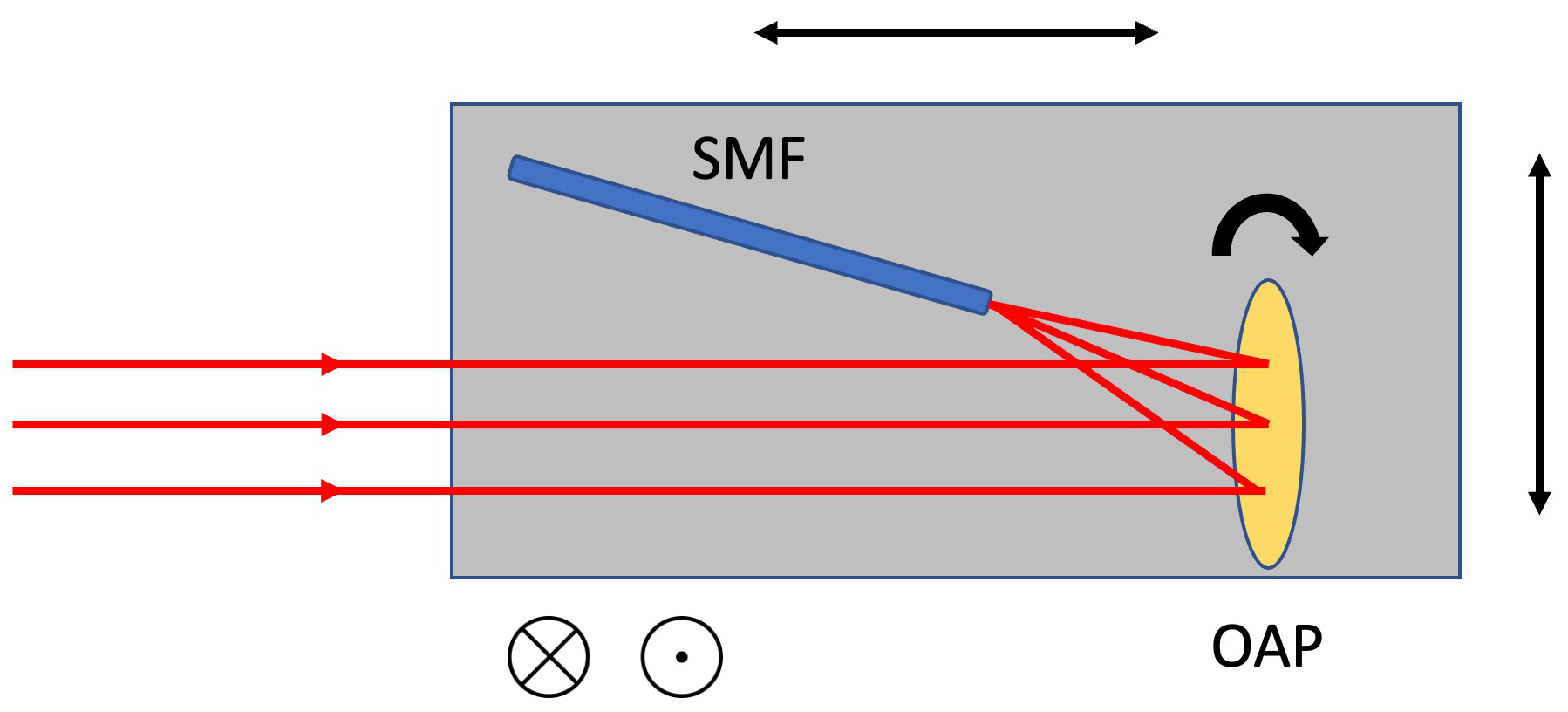}
    \caption{An alternative layout for the BIFROST fiber injection unit to that proposed in figure~\ref{fig:MYSTIC_style_injection}. Here the DDL and FSM are removed increasing the throughput by $\approx$~5\%. However the common block the OAP and SMF are mounted to must now move forwards and backwards to act as the DDL and in the plane of the beam to maximise coupling of the beam into the SMF.}
    \label{fig:simple_style_injection}
\end{figure}

An alternative layout is shown in figure~\ref{fig:simple_style_injection}. Here the dedicated DDL and FSM mirrors are removed and instead the OAP and SMF common mount (represented by the grey box in figure~\ref{fig:simple_style_injection}) is motorised. 

This plate is then motorised in all three axes, along the direction of the incoming beams to add additional path length, and within the plane of the incoming beam to centre the OAP on the beam. The advantage of this scheme is that it requires fewer optics, improving the throughput and reducing the wavefront error. The disadvantage this scheme is that the OAP fiber unit must move which will change the fibers shape. As an example the MIRC-X DDLs have a range of \SI{14}{\milli\meter} (Anugu et al. 2020)\cite{2020AJ....160..158A}. More work is required to verify if moving the fibers by such distances could alter the differential phase of orthogonal polarisation states (which would need a subsequent correction by the LNP, section~\ref{sec:Lithium_niobate_plates}) or even change the optical path length through the fibers.  

\section{Beam combination unit}

Once the beams of starlight exit the single mode optical fibers they will need to be interfered with each other via the beam combiner, to produce interference fringes on all six baselines for our four telescope combiner.

We are considering two different beam combination schemes, the AIO which has been utilised by a number of current and planned combiners (Monnier et al. 2018; Anugu et al. 2020; Mortimer et al. 2020)\cite{2018SPIE10701E..22M,2020SPIE11446E..0NA,2020SPIE11446E..0VM} as well as IO which is also utilised by a number of combiners (Le Bouquin et al. 2011; Gravity Collaboration et al. 2017; Pannetier et al. 2020)\cite{2011A&A...535A..67L,2017A&A...602A..94G,2020SPIE11446E..0TP}.

In the following sections we present a comparison of the two combination methods against different metrics such as sensitivity and the practicalities of implementing and operating each system. 

\subsection{Sensitivity}

Defining sensitivity by estimating a Signal to Noise Ratio (SNR) is not trivial in optical interferometry as it depends on many factors such as the number of photons reaching the detector, the visibility of the interference fringes, the number of telescopes being interfered and detector read noise. Here we calculate the ratio of the SNR for an AIO and IO four telescope combiner in two regimes, observations of very bright targets where we are photon noise dominated, and very faint targets where we are read noise dominated.

Our SNR calculations are based on the equation presented in Mourard et al (2017)\cite{2017JOSAA..34A..37M} which itself is derived from Gordon \& Buscher (2012)\cite{2012A&A...541A..46G} and is given by 

\begin{equation} \label{equ:powerspectrum_SNR}
    \textrm{SNR} = \dfrac{\left(\dfrac{VF_{0}}{N_{\textrm{tel}}}\right)^2}{\sqrt{\textrm{PhotonNoise} + \textrm{ReadNoise} + \textrm{CoupledTerms}}},
\end{equation}
where $\textrm{PhotonNoise}$, $\textrm{ReadNoise}$ represent the variance due to photon noise and read noise respectively and $\textrm{CoupledTerms}$ gives coupling terms between the two noise sources. The $\textrm{PhotonNoise}$ term is given by 

\begin{equation}
    \textrm{PhotonNoise} = 2N_{\textrm{ph}}\left(\dfrac{VF_{0}}{N_{\textrm{tel}}}\right)^2 + N_{\textrm{ph}}^2,
\end{equation}
the $\textrm{ReadNoise}$ term by

\begin{equation}
    \textrm{ReadNoise} = N_{\textrm{pix}}\sigma^2 + N_{\textrm{pix}}^2\sigma^4,
\end{equation}
the $\textrm{CoupledTerms}$ by

\begin{equation}
    \textrm{CoupledTerms} = 2\left(\dfrac{VF_{0}}{N_{\textrm{tel}}}\right)^2N_{\textrm{pix}}\sigma^2 + 2N_{\textrm{ph}}N_{\textrm{pix}}\sigma^2.
\end{equation}

Where $V$ is the measured visibility of the interference fringes, $F_0$ the number of stellar photons, $N_{\textrm{tel}}$ the number of telescopes being combined, $N_{\textrm{ph}}$ is the total number of photons recorded in the interferogram (from the source, sky background and thermal photons), $N_{\textrm{pix}}$ the number of pixels the fringes were sampled over and $\sigma$ the read noise of each pixel.

\subsubsection{Photon noise dominated} \label{sec:photon_noise_dom}

Following the methodology outlined in Mortimer (2021)\cite{Mortimer_PhD_thesis}, under the assumption that photon noise is the dominant source of noise (as is the case for very bright targets) and that the only source of photons are stellar photons (i.e. $N_{\textrm{ph}} = F_0$) the SNR in equation \ref{equ:powerspectrum_SNR} can be significantly simplified to 

\begin{equation}
    \textrm{SNR} = \dfrac{V\sqrt{F_0}}{\sqrt{2}N_{tel}}.
\end{equation}

It is then possible to take the ratio of the SNR for the AIO and IO beam combination schemes to calculate the relative signal to noise ratios, given by

\begin{equation} \label{eq:snr_photon_noise}
    \dfrac{\textrm{SNR}_{\textrm{AIO}}}{\textrm{SNR}_{\textrm{IO}}} = \sqrt{\dfrac{F_{0_{\textrm{AIO}}}}{F_{0_{\textrm{IO}}}}}\dfrac{V_{\textrm{AIO}}N_{tel_{\textrm{IO}}}}{V_{IO}N_{tel_{\textrm{AIO}}}}.
\end{equation}

The first step is to estimate the relative number of stellar photons that are detected by comparing the throughput of the AIO and IO combination schemes ($F_{0_{\textrm{AIO}}}/F_{0_{\textrm{IO}}}$). A typical AIO combiner optimised for sensitivity can reach an optical throughput of $\sim$ 80\% (Mortimer. 2021)\cite{Mortimer_PhD_thesis} whereas an IO combiner can reach around 65\% (Benisty et al. 2009)\cite{2009A&A...498..601B}. There are a couple of additional caveats to these values however, firstly for the AIO combiner, as all the beams are combined into the same fringe packet we will implement photometric channels similar to MIRC-X (Anugu et al. 2020)\cite{2020AJ....160..158A} which takes 20\% of the light, meaning the amount of light left in the interferogram is around 64\%. The 65\% for the IO chip is the sum of the flux at all outputs, however as the combiner is pairwise (each interferogram only combines one baseline at a time) the amount of light in each interferogram is only 1/3 the total amount of light from each input as it is split evenly between the three baselines each telescope goes on to form, meaning the throughput is around 22\% in each interferogram. Therefore $F_{0_{\textrm{AIO}}}/F_{0_{\textrm{IO}}} \approx 3$. 

The next term to consider is the instrumental fringe contrast, the visibility that would be recorded for an ideal point source. For IO combiners, PIONIER has been shown to have an instrumental contrast of 80\% in wideband observations, tending towards 100\% if spectral dispersion is used (Benisty et al. 2009)\cite{2009A&A...498..601B} and GRAVITY has demonstrated 95\% for polarised light (Perraut et al. 2018)\cite{2018A&A...614A..70P}, we therefore adopt an estimate of 95\%. 

In an AIO combiner the interference fringes are each sampled at different spatial frequencies so the signal from each baseline can be separated during data analysis, however the spatial frequencies have to be sufficiently separated to avoid crosstalk meaning that for a given number of telescopes the fringes have to span a certain range of spatial frequencies (Mortimer \& Buscher 2022)\cite{2022MNRAS.511.4619M}. This leads to competing requirements that one wishes to minimise the number of pixels readout in the interferogram to minimise read noise, but also wants to sample the highest spatial frequency fringes well enough that there is not a sufficient loss in contrast due to poor pixel sampling. A compromise is to sample the highest frequency fringes on three pixels per fringe cycle. The instrumental contrast in this case can then be estimated by $V = \textrm{sinc}(\pi f)$ where $f$ is the fringe frequency in cycles/pixel. For three pixels/cycle this gives a value of $V$ = 82\% which we adopt here.  

Finally $N_{tel_{\textrm{AIO}}}$ = 4 as all four telescopes are combined simultaneously whereas $N_{tel_{\textrm{IO}}}$ = 2 as it is a pairwise combiner. 

Entering the above values into equation~\ref{eq:snr_photon_noise} we estimate $\textrm{SNR}_{\textrm{AIO}}/\textrm{SNR}_{\textrm{IO}}$ = 0.71 in this photon noise dominated regime giving IO a marginal but not insignificant advantage over AIO despite the three times higher throughput of AIO. This is mainly due to the pairwise combination scheme which the SNR goes linearly with, compared to the throughput ratio which the SNR only goes with the square root of. 

\subsubsection{Read noise dominated} \label{sec:read_noise_dominated}

Read noise in detectors used for optical interferometry has improved dramatically in recent years, from around 10 e-/pixel (Le Bouquin et al. 2011)\cite{2011A&A...535A..67L} to around 0.3 e-/pixel with the SAPHIRA detector (Lanthermann et al. 2019)\cite{2019A&A...625A..38L}. However, as Mortimer (2021)\cite{Mortimer_PhD_thesis} showed, readnoise can still be significant for faint targets. The simplified SNR equation in this case is found by assuming that only readnoise contributes to the noise and so the SNR is given by 

\begin{equation}
    \textrm{SNR} = \dfrac{\left(\dfrac{VF_{0}}{N_{\textrm{tel}}}\right)^2}{\sqrt{N_{\textrm{pix}}\sigma^2 + N_{\textrm{pix}}^2\sigma^4}}.
\end{equation}

Again taking the ratio of the SNR for both the AIO and IO combination schemes and rearranging we arrive at 

\begin{equation}
    \dfrac{\textrm{SNR}_{\textrm{AIO}}}{\textrm{SNR}_{\textrm{IO}}} = \left(\dfrac{V_{\textrm{AIO}}F_{0_{\textrm{AIO}}}N_{\textrm{Tel}_{\textrm{IO}}}}{V_{\textrm{IO}}F_{0_{\textrm{IO}}}N_{\textrm{Tel}_{\textrm{AIO}}}}\right)^2 \sqrt{\dfrac{N_{\textrm{pix}_{\textrm{IO}}}\sigma^2 + N_{\textrm{pix}_{\textrm{IO}}}^2\sigma^4}{N_{\textrm{pix}_{\textrm{AIO}}}\sigma^2 + N_{\textrm{pix}_{\textrm{AIO}}}^2\sigma^4}}.
\end{equation}

The value for all the terms have been defined already in section~\ref{sec:photon_noise_dom} except for the number of pixels the interferogram is readout over. For IO, in theory only four pixels need to be read out to sample the four outputs per baseline, however we assume this is raised to eight pixels (two per output) to minimise the risk posed to the instrument if the light happens to land on a bad pixel. 

\begin{figure}
    \centering
    \includegraphics[width=0.9\columnwidth]{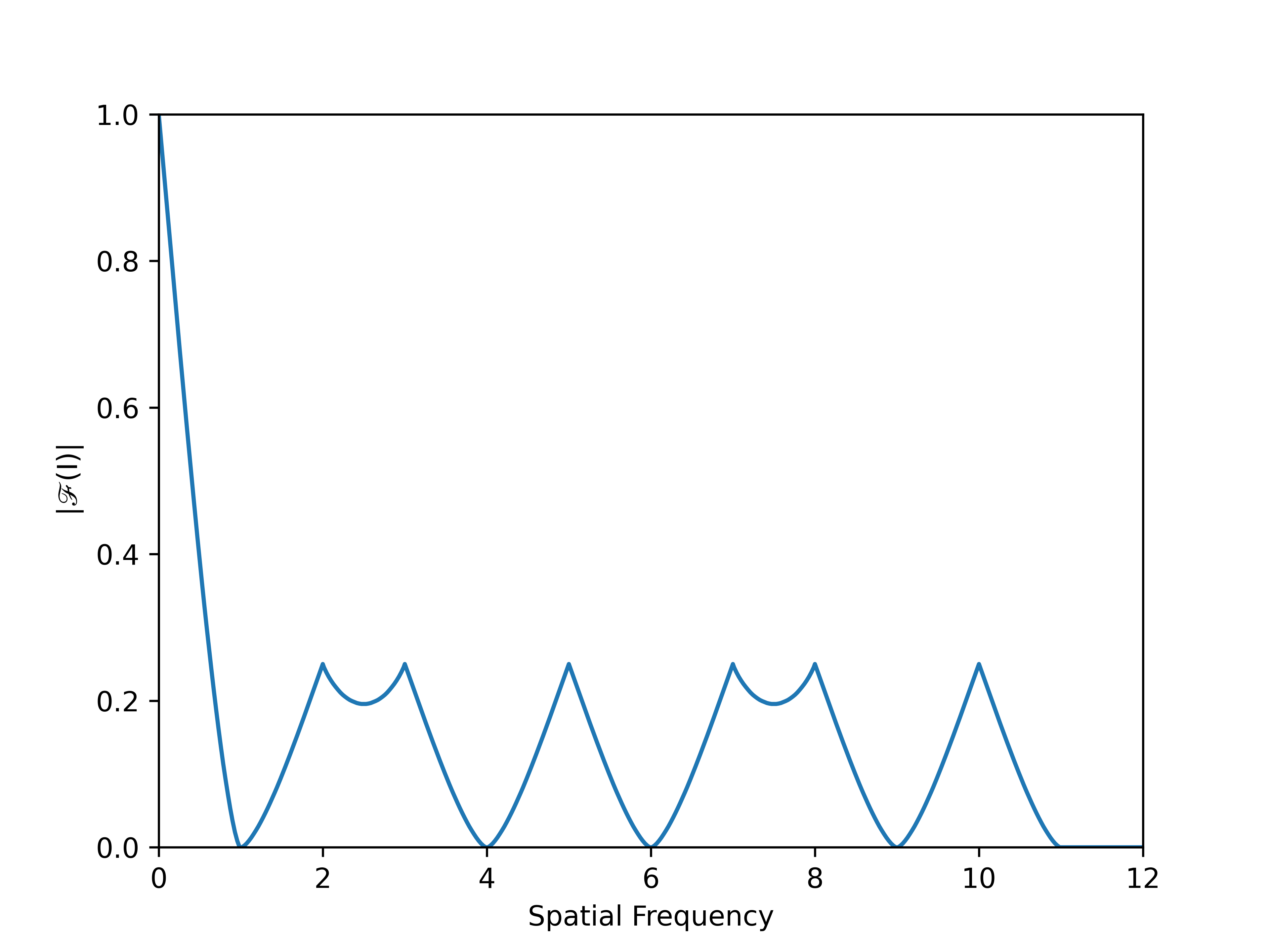}
    \caption{The most compact power spectrum which minimises crosstalk for a four telescope AIO combiner as per the methodology of Mortimer \& Buscher (2022)\cite{2022MNRAS.511.4619M}. As the figure shows the highest spatial frequency interference fringes are at five times higher a spatial frequency than the lowest. Taking this and the conditions described in the text gives a pixels sampling of 60 pixels across the AIO interferogram.}
    \label{fig:BIFROST_power_spectrum}
\end{figure}

The number of pixels readout for an AIO combiner can be significantly more. We start by stating that the condition for sampling our interference fringes are that we wish to sample at least three pixels/cycle on the highest frequency fringes and at least four cycles of the interference fringes for the lowest spatial frequency fringes (as is the case for MIRC-X (Anugu et al. 2020)\cite{2020AJ....160..158A}). We can relate the conditions by looking at the ideal non-redundant spacing for a four telescope interferometer. Following the methodology discussed in Mortimer \& Buscher 2022)\cite{2022MNRAS.511.4619M} the best beam spacing that minimises the range of spatial frequencies while also minimising crosstalk in the power spectrum shown in figure \ref{fig:BIFROST_power_spectrum} which shows the highest spatial frequency fringes are five times higher than the lowest. Therefore if three pixels/cycle are sampled on the highest spatial frequency fringes then 15 pixels/cycle will be sampled on the lowest, multiplying this by four to sample four cycles results in 60 pixels being readout in the AIO scheme. It is worth noting that this beam spacing is not necessarily the configuration that would be used in practise but instead represents the most compact (hence requiring the fewest pixels readout) layout that could be used while still avoiding the worse effects of crosstalk. 

This then gives the value of $\textrm{SNR}_{\textrm{AIO}}/\textrm{SNR}_{\textrm{IO}}$ = 0.29, making the IO combination scheme approximately 3.5$\times$ more sensitive in this readnoise dominated regime. 

\subsection{Implementation}

In addition to sensitivity considerations the actual implementation of each beam combination scheme must be considered as both schemes have their advantages and disadvantages. 

With regards to procurement AIO combiners can be cheaper and faster as they can be built from mostly off-the-shelf optical components. This removes any design or tooling costs involved in making low production run custom optical components and they can be delivered much faster as there is no development lead time. The opposite is true for IO combiners where due to the low volume and unique use case each IO chip to date has been designed and fabricated specifically for each combiner. 

Considering the practical implementation of such systems IO offers the advantage that once the chip has been fabricated the alignment should be much more stable as it is a monolithic unit whereas the AIO system requires many individual optics to be carefully aligned with respect to each other to work. This advantage of IO could be significant for BIFROST, which is to be a visitor instrument at the VLTI, as the alignment of the instrument will not be maintained between observing runs which could lead to a significant overhead if the instrument has to be realigned at the start of each observing run. 

Another advantage of IO to consider is it's compact size, a 4 telescope IO chip is of order a few centimetres in size (Benisty et al. 2009)\cite{2009A&A...498..601B} compared to of order \SI{50}{\centi\meter} for an AIO combiner (Mortimer. 2021)\cite{Mortimer_PhD_thesis}. This reduced footprint could be significant for BIFROST which will share one optical table with four other interferometric instruments as part of the Asgard Suite. 

The final consideration is the chromaticity of the combiners. AIO offers the advantage that it can be made from mostly reflective components making it fairly achromatic with combiners having been built that are capable of operating over broad wavelength ranges such as $\lambda$~=~\SI{1.1}{} - \SI{2.4}{\micro\meter} (Mortimer. 2021)\cite{Mortimer_PhD_thesis}. IO combiners on the other hand are chromatic, early discussions with vendors suggests that BIFROST will require two IO combiners, one designed to operate in the Y+J band ($\lambda$ = \SI{1.0}{}~-~\SI{1.4}{\micro\meter}) and one in the H band ($\lambda$ = \SI{1.5}{}~-~\SI{1.7}{\micro\meter}). Such a two chip system will add cost and complexity to the design of BIFROST.


\section{Future work}

In this proceeding we have outlined the critical subsystems required for the operation of BIFROST at the VLTI as well as presented a comparison between the AIO and IO beam combination schemes. The next steps in the development of the instrument will be to design the critical subsystems and progress towards a preliminary design review of BIFROST in Q1 of 2023. This will be followed by the final design review and integration of the instrument at the university of Exeter before commissioning at the VLTI in late 2024. 

\acknowledgments 

We acknowledge support from ERC Consolidator Grant (Grant Agreement ID 101003096) and STFC Consolidated Grant (ST/V000721/1).

\bibliography{report} 
\bibliographystyle{spiebib} 

\end{document}